 \documentclass[UTF8,final,3p,times,twocolumn]{elsarticle}
\usepackage{amssymb}

\journal{Physics Letters B}

\begin{document}

\begin{frontmatter}

%% Title, authors and addresses

%% use the tnoteref command within \title for footnotes;
%% use the tnotetext command for theassociated footnote;
%% use the fnref command within \author or \address for footnotes;
%% use the fntext command for theassociated footnote;
%% use the corref command within \author for corresponding author footnotes;
%% use the cortext command for theassociated footnote;
%% use the ead command for the email address,
%% and the form \ead[url] for the home page:
%% \title{Title\tnoteref{label1}}
%% \tnotetext[label1]{}
%% \author{Name\corref{cor1}\fnref{label2}}
%% \ead{email address}
%% \ead[url]{home page}
%% \fntext[label2]{}
%% \cortext[cor1]{}
%% \address{Address\fnref{label3}}
%% \fntext[label3]{}

\title{Experimentally well-constrained masses of $^{27}$P and $^{27}$S: Implications for studies of explosive binary systems}

%% use optional labels to link authors explicitly to addresses:
%% \author[label1,label2]{}
%% \address[label1]{}
%% \address[label2]{}
\author[CIAE,HKU,SJTU]{L.~J.~Sun\corref{cor1}}
\author[CIAE,HKU,IMP]{X.~X.~Xu\corref{cor1}\corref{cor2}}
\ead{xinxing@impcas.ac.cn}
\author[IMP,NuGrid]{S.~Q.~Hou\corref{cor1}}
\author[CIAE,GNU]{C.~J.~Lin\corref{cor2}}
\ead{cjlin@ciae.ac.cn}
\author[UPC,IEEC]{J.~Jos\'{e}\corref{cor2}}
\ead{jordi.jose@upc.edu}
\author[HKU]{J.~Lee\corref{cor2}}
\ead{jleehc@hku.hk}
\author[BNU,UCAS]{J.~J.~He}
\author[PKU]{Z.~H.~Li}
\author[IMP]{J.~S.~Wang}
\author[SYSU]{C.~X.~Yuan}
\author[UVic,JINA,NuGrid]{F.~Herwig}
\author[Hull,NuGrid]{J.~Keegans}
\author[MSU,NSCL]{T.~Budner}
\author[CIAE]{D.~X.~Wang}
\author[PKU]{H.~Y.~Wu}
\author[HKU]{P.~F.~Liang}
\author[IMP]{Y.~Y.~Yang}
\author[IMP]{Y.~H.~Lam}
\author[IMP]{P.~Ma}
\author[LZU,IMP]{F.~F.~Duan}
\author[IMP,LZU]{Z.~H.~Gao}
\author[IMP]{Q.~Hu}
\author[IMP]{Z.~Bai}
\author[IMP]{J.~B.~Ma}
\author[IMP]{J.~G.~Wang}
\author[GNU,CIAE]{F.~P.~Zhong}
\author[PKU]{C.~G.~Wu}
\author[PKU]{D.~W.~Luo}
\author[PKU]{Y.~Jiang}
\author[PKU]{Y.~Liu}
\author[IMP,UCAS]{D.~S.~Hou}
\author[IMP,UCAS]{R.~Li}
\author[CIAE]{N.~R.~Ma}
\author[IMP,FDU]{W.~H.~Ma}
\author[IMP]{G.~Z.~Shi}
\author[IMP]{G.~M.~Yu}
\author[IMP]{D.~Patel}
\author[IMP,UCAS]{S.~Y.~Jin}
\author[YNU,IMP]{Y.~F.~Wang}
\author[YNU,IMP]{Y.~C.~Yu}
\author[SWU,IMP]{Q.~W.~Zhou}
\author[SWU,IMP]{P.~Wang}
\author[HEU]{L.~Y.~Hu}
\author[PKU]{X.~Wang}
\author[PKU]{H.~L.~Zang}
\author[HKU]{P.~J.~Li}
\author[HKU]{Q.~Q.~Zhao}
\author[CIAE]{L.~Yang}
\author[CIAE]{P.~W.~Wen}
\author[CIAE]{F.~Yang}
\author[CIAE]{H.~M.~Jia}
\author[BUAA]{G.~L.~Zhang}
\author[BUAA,CIAE]{M.~Pan}
\author[BUAA]{X.~Y.~Wang}
\author[CIAE]{H.~H.~Sun}
\author[IMP]{Z.~G.~Hu}
\author[IMP]{R.~F.~Chen}
\author[IMP]{M.~L.~Liu}
\author[IMP]{W.~Q.~Yang}
\author[SJTU]{Y.~M.~Zhao}
\author[CIAE]{H.~Q.~Zhang}

\cortext[cor1]{These authors contributed equally to this work and should be considered as co-first authors.}
\cortext[cor2]{Corresponding authors}
\address[CIAE]{Department of Nuclear Physics, China Institute of Atomic Energy, Beijing 102413, China}
\address[HKU]{Department of Physics, The University of Hong Kong, Hong Kong, China}
\address[SJTU]{School of Physics and Astronomy, Shanghai Jiao Tong University, Shanghai 200240, China}
\address[IMP]{Institute of Modern Physics, Chinese Academy of Sciences, Lanzhou 730000, China}
\address[GNU]{College of Physics and Technology, Guangxi Normal University, Guilin 541004, China}
\address[UPC]{Departament de F\'{i}sica, EEBE, Universitat Polit\'{e}cnica de Catalunya, Av./ Eduard Maristany 10, E-08930 Barcelona, Spain}
\address[IEEC]{Institut d’Estudis Espacials de Catalunya (IEEC), Ed. Nexus-201, C/ Gran Capit\'{a} 2-4, E-08034 Barcelona, Spain}
\address[BNU]{Key Laboratory of Beam Technology of Ministry of Education, College of Nuclear Science and Technology, Beijing Normal University, Beijing 100875, China}
\address[UCAS]{University of Chinese Academy of Sciences, Beijing 100049, China}
\address[PKU]{State Key Laboratory of Nuclear Physics and Technology, School of Physics, Peking University, Beijing 100871, China}
\address[SYSU]{Sino-French Institute of Nuclear Engineering and Technology, Sun Yat-Sen University, Zhuhai 519082, China}
\address[UVic]{Department of Physics \& Astronomy, University of Victoria, Victoria, BC, V8W 2Y2, Canada}
\address[JINA]{Joint Institute for Nuclear Astrophysics, Center for the Evolution of the Elements, Michigan State University, East Lansing, MI 48824, USA}
\address[Hull]{E.A. Milne Centre for Astrophysics, Department of Physics \& Mathematics, University of Hull, Hull HU6 7RX, UK}
\address[MSU]{Department of Physics and Astronomy, Michigan State University, East Lansing, Michigan 48824, USA}
\address[NSCL]{National Superconducting Cyclotron Laboratory, Michigan State University, East Lansing, Michigan 48824, USA}
\address[LZU]{School of Nuclear Science and Technology, Lanzhou University, Lanzhou 730000, China}
\address[FDU]{Institute of Modern Physics, Fudan University, Shanghai 200433, China}
\address[YNU]{School of Physics and Astronomy, Yunnan University, Kunming 650091, China}
\address[SWU]{School of Physical Science and Technology, Southwest University, Chongqing 400044, China}
\address[HEU]{Fundamental Science on Nuclear Safety and Simulation Technology Laboratory, Harbin Engineering University, Harbin 150001, China}
\address[BUAA]{School of Physics and Nuclear Energy Engineering, Beihang University, Beijing 100191, China}
\address[NuGrid]{NuGrid collaboration, http://www.nugridstars.org}
\address[RIBLL]{RIBLL Collaboration}

\begin{abstract}
The mass of $^{27}$P was predicted to impact the X-ray burst (XRB) model predictions of burst light curves and the composition of the burst ashes. To address the uncertainties and inconsistencies in the reported $^{27}$P masses in literature, a wealth of information has been extracted from the $\beta$-decay spectroscopy of the drip-line nucleus $^{27}$S. We determine the most precise mass excess of $^{27}$P to date to be $-659(9)$~keV, which is 63~keV (2.3$\sigma$) higher than the AME2016 recommended value of $-722(26)$~keV. The experimentally unknown mass excess of $^{27}$S was estimated to be 17030(400)~keV in AME2016, and we constrain this mass to be 17678(77)~keV based on the measured $\beta$-delayed two-proton energy. In the temperature region of $(0.06-0.3)$~GK, the $^{26}$Si$(p,\gamma)^{27}$P reaction rate determined in this work is significantly lower than the rate recommended in the reaction rate libraries, up to two orders of magnitude around 0.1~GK. The impact of these newly determined masses and well-constrained rate on the modeling of the explosive astrophysical scenarios has been explored by hydrodynamic nova and post-processing XRB models. No substantial change was found in the nova contribution to the synthesis of galactic $^{26}$Al or in the XRB energy generation rate, but we found that the calculated abundances of $^{26}$Al and $^{26}$Si at the last stage of XRB are increased by a factor of 2.4. We also conclude that $^{27}$S is not a significant waiting point in the rapid proton capture process.
%The experimentally unknown mass excess of $^{27}$S was estimated to be 17030(400)~keV in AME2016, and we better constrain this mass to be 17678(77)~keV based on the measured $\beta$-delayed two-proton energy.
\end{abstract}

\begin{keyword}
%% keywords here, in the form: keyword \sep keyword
$\beta$-delayed proton emission \sep $\beta$-delayed $\gamma$-ray emission \sep Masses of $^{27}$P and $^{27}$S \sep X-ray bursts \sep Nova outbursts
%% PACS codes here, in the form: \PACS code \sep code
%\PACS 07.77.-n \sep 29.40.Wk
%% MSC codes here, in the form: \MSC code \sep code
%% or \MSC[2008] code \sep code (2000 is the default)
\end{keyword}

\end{frontmatter}

\section{Introduction}
Type I X-ray bursts (XRB) and classical novae are the two most frequent types of thermonuclear stellar explosions in the Galaxy. They are powered by thermonuclear runaways occurring in the accreted envelopes of compact objects in stellar binary systems. In the case of XRBs, hydrogen- or helium-rich material is transferred from a low mass main sequence or red giant star onto the surface of a neutron star, while nova explosions occur in a similar system with a white dwarf in place of the neutron star. As they are driven by a suite of nuclear processes, accurate nuclear physics inputs such as $\beta$ decay rates, masses, and nuclear reaction rates of neutron-deficient rare isotopes are needed to model the energy production and nucleosynthesis in these explosions. Our understanding of these systems has greatly improved with time, but despite decades of work, many open questions remain~\cite{Jose_2016,Parikh_AIPA2014,Schatz_NPA2006,Jose_NPA2006,Jose_JPG2007,Jose_APJ1998}.

A recent systematic investigation of the impact of nuclear mass uncertainties on XRB models found that the mass uncertainties of $^{27}$P can strongly affect the model predictions of the burst light curve and the composition of the burst ashes in a typical mixed H/He burst~\cite{Schatz_APJ2017}. A recent $^{27}$S $\beta$-decay measurement using an optical time projection chamber~\cite{Janiak_PRC2017} reported a mass excess of $\mathrm{\Delta}(^{27}$P$)=-640(30)$~keV, which is inconsistent with the mass excess of $\mathrm{\Delta}(^{27}$P$)=-716(7)$~keV calculated using the isobaric mass multiplet equation~\cite{Schatz_APJ2017} and $\mathrm{\Delta}(^{27}$P$)=-722(26)$~keV reported by AME2016~\cite{Wang_CPC2017}. The AME2016 value was the weighted average of two previously measured mass excesses of $-753(35)$~keV~\cite{Benenson_PRC1977} and $-670(41)$~keV~\cite{Caggiano_PRC2001}, which are also somewhat inconsistent. The most recent result of $\mathrm{\Delta}(^{27}$P$)=-685(42)$ measured via isochronous mass spectrometry in the Cooler Storage Ring~\cite{Fu_PRC2018} did not resolve the existing discrepancies. Additionally, $^{27}$S was considered a nuclear waiting-point in the thermonuclear reaction network, and its mass uncertainty was thought to impact the nucleosynthesis in some XRB models~\cite{Parikh_PRC2009,Parikh_PPNP2013}. Nevertheless, the mass excess of $^{27}$S is unknown experimentally, and AME2016 roughly estimated the mass to be $\mathrm{\Delta}(^{27}$S$)=17030(400)$~keV~\cite{Wang_CPC2017}. Hence, experimental efforts should be made to better quantify the mass excesses of $^{27}$P and $^{27}$S.

Furthermore, the origin of large amounts of $^{26}$Al in the interstellar medium of the galaxy has been a focus of interdisciplinary investigations from astronomy, astrophysics, and nuclear physics~\cite{Diehl_N2006}. The nova nucleosynthesis of $^{26}$Al is dominated by a reaction sequence of $^{24}$Mg$(p,\gamma$)$^{25}$Al$(\beta^+)^{25}$Mg$(p,\gamma)^{26}$Al$(p,\gamma)^{27}$Si, but this sequence may be bypassed through $^{25}$Al($p,\gamma)^{26}$Si$(p,\gamma)^{27}$P~\cite{Prantzos_PR1996,Jose_APJ1997}. Under a wide temperature range of $(0.1-2)$~GK, the $^{26}$Si$(p,\gamma)^{27}$P reaction rate was found to be dominated only by a single resonant proton capture on the $^{26}$Si ground state to the $3/2^+$ first excited state in $^{27}$P. According to previous nova nucleosynthesis calculations~\cite{Jose_APJ1999} the $^{26}$Si$(p,\gamma)^{27}$P rate was not expected to play a critical role, but it should be noted that a complete experimental constraint on the thermonuclear $^{26}$Si$(p,\gamma)^{27}$P rate had never been set. Estimates of those key resonance strengths have relied on limited experimental information on the structure of $^{27}$P, supplemented by shell model calculations or the mirror nucleus information~\cite{Caggiano_PRC2001,Moon_NPA2005,Jung_PRC2012,Togano_PRC2011,Marganiec_PRC2016,Herndl_PRC1995,Rauscher_PRC1997,Guo_PRC2006,Timofeyuk_PRC2008,Qi_SCPMA2009,Fortune_PRC2015}. A reevaluation of the role of the $^{26}$Si$(p,\gamma)^{27}$P reaction with a new $^{27}$P mass or $Q$-value may benefit the long-standing study of the galactic $^{26}$Al origin.

In this Letter, we report high-statistics $\beta$-decay spectroscopy of $^{27}$S. The emitted particles and $\gamma$ rays were measured simultaneously for the first time for the $\beta$ decay of $^{27}$S, allowing us to determine an accurate $^{27}$P mass excess and to place an experimental constraint on the $^{27}$S mass excess. Additionally, we provide a new resonance energy and strength for the key $3/2^+$ resonance in the $^{26}$Si$(p,\gamma)^{27}$P reaction, enabling a reliable reexamination of impact of these masses and the reaction rate on XRB and nova models.

%\textit{Experiment}
\section{Experiment}
The experiment was carried out at the Radioactive Ion Beam Line of the Heavy Ion Research Facility in Lanzhou (HIRFL-RIBLL)~\cite{Zhan_NPA2008,Sun_NIMA2003}. A 80.6-MeV/nucleon, $\sim$5.4-$p$nA $^{32}$S$^{16+}$ primary beam was produced using the K69 Sector Focus Cyclotron and the K450 Separate Sector Cyclotron. The secondary $^{27}$S ions were produced via the projectile fragmentation of the $^{32}$S beam impinging on a 1581-$\mathrm{\mu}$m thick $^9$Be target. The charged-particle detection system was composed of several double-sided silicon strip detectors (DSSD)~\cite{Xu_NST2018} and quadrant silicon detectors~\cite{Bao_CPC2014}, which has been successfully employed in our previous $\beta$-decay experiments~\cite{Sun_CPL2015,Sun_PRC2017,Xu_PLB2017,Wang_IJMPE2018,Wang_NST2018,Wang_EPJA2018,Wang_PLB2018}. The $^{27}$S ions were implanted into DSSD1~(142~$\mathrm{\mu}$m), DSSD2~(40~$\mathrm{\mu}$m), and DSSD3~(304~$\mathrm{\mu}$m) with proportions of 0.6\%, 40.6\%, and 58.7\%, respectively. The implantation and subsequent decay were then correlated in space and time. DSSD2 is the thinnest W1-type DSSD ever produced by the Micron Semiconductor Ltd, which aimed to detect low-energy protons with a minimal $\beta$-summing. DSSD3 has a higher detection efficiency for high-energy protons and $\beta$ particles, being an important supplement to DSSD2. The implanted ions can be identified on an event-by-event basis over the entire experiment, and thus, an accurate number of implanted $^{27}$S ions, as well as the absolute proton and $\gamma$-ray intensities, can be determined accordingly. In addition, five clover-type high-purity germanium (HPGe) detectors were employed to measure the $\gamma$ rays~\cite{Sun_NIMA2015}. The present data set and analysis procedures have been detailed in Ref.~\cite{Sun_PRC2019}.
%In this experiment, a total of $4.7\times10^4$ $^{27}$S ions was implanted, which is much higher than the statistics reported in previous $^{27}$S decay studies, $\sim$$1\times10^4$ in Ref.~\cite{Canchel_EPJA2001} and 1267 in Ref.~\cite{Janiak_PRC2017}. A 1546~$\mathrm{\mu}$m thick QSD1 was installed downstream to detect the emitted $\beta$ particles and two $\sim$300~$\mathrm{\mu}$m thick QSD2 and QSD3 were installed at the end to veto the possible contaminants from $^1$H, $^2$H, $^3$H, and $^4$He coming along with the beam. Both silicon detectors and preamplifiers were cooled by using a cryogenic system, which dramatically reduced the leakage current of detectors, and enabled us to achieve a better resolution and to maintain the operation stability of the detection system
%
\section{Results}
%\subsection{Half-life}
%A new half-life of $^{27}$S was determined to be 16.3(2)~ms by fitting the time differences between implantation-decay events with a function composed of an exponential decay and a constant background component. The uncertainty was directly derived from the fitting program, in which the half-life was treated as a free parameter and no preset parameters were involved. The present half-life is in good agreement with the literature values of 21(4)~ms~\cite{Borrel_NPA1991}, 15.5(15)~ms~\cite{Canchel_EPJA2001}, and 15.5(16)~ms~\cite{Janiak_PRC2017}, but with much higher accuracy owing to the high statistics and the large time-correlation window between implantation-decay events achieved in this work. It is worthwhile to improve the precision of the lifetime of $^{27}$S, which is a waiting point nucleus in XRBs nucleosynthesis~\cite{Parikh_PPNP2013}.

%\textit{Decay properties}
\subsection{Decay properties}
\begin{figure}
\begin{center}
\includegraphics[width=7.5cm]{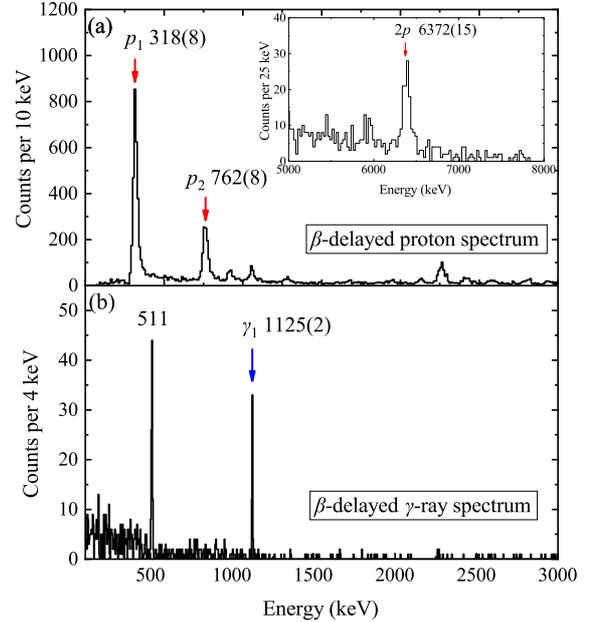}
\caption{\label{Spectra}(a) Low-energy region of the $\beta$-delayed proton spectrum measured by DSSD2 and DSSD3 and high-energy region demonstrating a $\beta$-delayed two-proton peak from $^{27}$S decay (inset). (b) Cumulative $\gamma$-ray spectrum measured by the HPGe detectors in coincidence with the $\beta$ particles from $^{27}$S decay measured by DSSD3. The proton and $\gamma$-ray peaks of interest are labeled with their center-of-mass energies given in units of keV (with errors indicated in the parentheses).}
\end{center}
\end{figure}

Figure~\ref{Spectra}(a) shows the cumulative $\beta$-delayed proton spectrum from $^{27}$S decay measured by DSSD2 and DSSD3. This spectrum represents the highest-statistics proton spectrum resulting from $^{27}$S decay to date, in which the center-of-mass energies and intensities of the two lowest-energy $\beta$-delayed proton peaks are determined to be $E_{p1}=318(8)$~keV, $E_{p2}=762(8)$~keV, $I_{p1}=23.1(21)\%$, and $I_{p2}=8.9(10)\%$, respectively, using the calibration based on a previous measurement~\cite{Thomas_EPJA2004}. The two-proton emission from the $T=5/2$ isobaric analog state (IAS) in $^{27}$P to the $^{25}$Al ground state was identified in previous $^{27}$S decay studies~\cite{Borrel_NPA1991,Canchel_EPJA2001}, whereas the two measured center-of-mass energies, $E_{2p}=6410(45)$~keV and $E_{2p}=6270(50)$~keV, were mutually inconsistent by $2.1\sigma$. As can be seen in the inset of Fig.~\ref{Spectra}(a), we measure this two-proton energy to be 6372(15)~keV, which falls between these two previous results~\cite{Borrel_NPA1991,Canchel_EPJA2001}. We also applied the relationship between the energy loss, position, and path length of the escaping particles in different DSSDs to confirm this is indeed a two-proton peak rather than an one-proton peak at the same energy~\cite{Xu_PLB2017}. Figure~\ref{Spectra}(b) shows the $\gamma$-ray spectrum measured by the HPGe detectors in coincidence with the $^{27}$S $\beta$-decay events in DSSD3. A $\beta$-delayed $\gamma$ ray with $E_{\gamma1}=1125(2)$~keV is clearly observed, which has been assigned as the deexcitation from the $3/2^+$ first excited state to the $1/2^+$ ground state in $^{27}$P~\cite{Gade_PRC2008}. The present energy is in excellent agreement with the values of 1125(6)~keV and 1119(8)~keV reported in a recent in-beam $\gamma$-ray spectroscopy~\cite{Longfellow_PRC2019}, and our precision is an improvement of a factor of 3. The absence of $\gamma$ rays other than the one at 1125~keV has also been confirmed by Ref.~\cite{Longfellow_PRC2019}. The absolute intensity per $^{27}$S $\beta$ decay is determined to be $I_{\gamma1}=31.1(86)\%$. Thus, the ratio of the $\gamma$-ray branch to proton branch for the $3/2^+$ first excited state of $^{27}$P $I_{\gamma}/I_p=1.35(39)$ is experimentally determined for the first time.%Here, the statistical and all the available systematic uncertainties have been taken into account for the results discussed above.

%An anticoincidence with $\beta$-particle signals in QSD1 efficiently suppresses the $\beta$-summing effect on the proton spectrum measured by the thicker DSSD3. Two low-energy $\beta$-delayed proton peaks from $^{27}$S decay are marked with $p_1$ and $p_2$, together with their respective center-of-mass energies of $E_{p1}=318(8)$~keV and $E_{p2}=762(8)$~keV. The energies are in good agreement with the previous values of $E_{p1}=332(30)$~keV and $E_{p2}=737(30)$~keV~\cite{Janiak_PRC2017}, and the precision has been improved by a factor of $\sim$4 in both cases. They are not observed in coincidence with $\gamma$ rays, and hence $p_1$ and $p_2$ should be assigned as the proton emissions from the $3/2^+$ first excited state and the $5/2^+$ second excited state of $^{27}$P, respectively, to the ground state of $^{26}$Si. The intensities for the two proton emissions per $^{27}$S decay are determined to be $I_{p1}=23.1(21)\%$ and $I_{p2}=8.9(10)\%$, consistent with the literature values~\cite{Janiak_PRC2017}.

\subsection{Mass of $^{27}$P}
\begin{figure}
\begin{center}
\includegraphics[width=7.5cm]{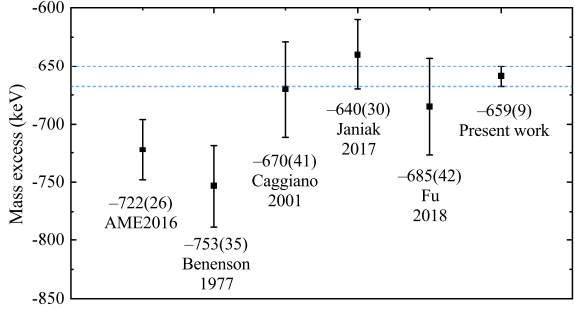}
\caption{\label{Mass}Mass excesses of $^{27}$P measured in the present work compared with the recommended value from AME2016~\cite{Wang_CPC2017} and values previously measured by Beneson \textit{et al}.~\cite{Benenson_PRC1977}, Caggiano \textit{et al}.~\cite{Caggiano_PRC2001}, Janiak \textit{et al}.~\cite{Janiak_PRC2017}, and Fu \textit{et al}.~\cite{Fu_PRC2018}.}
\end{center}
\end{figure}

By combining the excitation energy $E_{\gamma1}=1125(2)$~keV with the proton-decay energy of $E_{p1}=318(8)$~keV of the first excited state in $^{27}$P, the proton-separation energy of $^{27}$P is determined to be 807(9)~keV with the relation $S_{p}(^{27}$P$)=E_{\gamma1}-E_{p1}$. The present $S_{p}(^{27}$P) value is 63~keV (2.3$\sigma$) lower than the recommended value of $S_{p}(^{27}$P$)=870(26)$~keV from AME2016~\cite{Wang_CPC2017}. Our result is consistent with the $S_{p}(^{27}$P$)=788(30)$~keV reported by Ref.~\cite{Janiak_PRC2017} based on a $E_{p1}=332(30)$~keV and a $E_{\gamma1}=1120(8)$~keV~\cite{Gade_PRC2008}. Combined with the well-known mass excesses of $^{26}$Si and $^{1}$H from AME2016~\cite{Wang_CPC2017}, the mass excess of the $^{27}$P is determined to be $-659(9)$~keV via the relation $\mathrm{\Delta}(^{27}$P$)=\mathrm{\Delta}(^{26}$Si$)+\mathrm{\Delta}(^{1}$H$)-S_{p}(^{27}$P), which represents the most precise $^{27}$P mass measurement to date (see Fig.~\ref{Mass}), and the total uncertainty is dominated by the measured proton energy uncertainty. The present $\mathrm{\Delta}(^{27}$P) is 63~keV (2.3$\sigma$) higher than that of $\mathrm{\Delta}(^{27}$P$)=-722(26)$~keV recommended by AME2016~\cite{Wang_CPC2017} but is in good agreement with two recent measurements of $\mathrm{\Delta}(^{27}$P$)=-685(42)$~keV~\cite{Fu_PRC2018} and $\mathrm{\Delta}(^{27}$P$)=-640(30)$~keV~\cite{Janiak_PRC2017}.

%\textit{Mass of $^{27}$S}
\subsection{Mass of $^{27}$S}
Combining the energy of the measured two-proton emission $E_{2p}=6372(15)$~keV with the well-known mass excesses of $\mathrm{\Delta}(^{25}$Al$)=-8915.97(6)$~keV and $\mathrm{\Delta}(^{1}$H$)=7288.97061(9)$~keV from AME2016~\cite{Wang_CPC2017}, the mass excess of the $T=5/2$ IAS in $^{27}$P is determined to be 12034(15)~keV using the relation $\mathrm{\Delta}(^{27}$P~IAS$)=\mathrm{\Delta}(^{25}$Al$)+2\mathrm{\Delta}(^{1}$H$)+E_{2p}$. Combined with the aforementioned mass excess of the $^{27}$P ground state, the excitation energy of the $^{27}$P IAS is determined to be 12693(18)~keV, which compares fairly well with the excitation energy of 12677~keV calculated by the shell model~\cite{Shimizu_CPC2019,Yuan_PRC2014}. The mass excess of $^{27}$S is estimated to be $\mathrm{\Delta}(^{27}$S$)=17678(77)$~keV using the relation $\mathrm{\Delta}(^{27}$S$)=\mathrm{\Delta}(^{27}$P~IAS$)+\mathrm{\Delta}E_C-\mathrm{\Delta}_{n\mathrm{H}}$, where $\mathrm{\Delta}_{n\mathrm{H}}=782.3465(5)$~keV is the mass difference between the neutron and hydrogen atom. $\mathrm{\Delta}E_C=6426(76)$~keV is the Coulomb displacement energy calculated by using the semiempirical relation given by Ref.~\cite{Miernik_APPB2013} with the corresponding isospin of $T=5/2$, mean atomic number of $\overline{Z}=15.5$, and mass number of $A=27$ in this case. The present $\mathrm{\Delta}(^{27}$S$)$ is 648~keV(1.6$\sigma$) higher than the $\mathrm{\Delta}(^{27}$S$)=17030(400)$~keV estimated by AME2016 and correspondingly reduces the $S_{p}(^{27}$S$)=581(215)$~keV compared with the AME2016 value of $S_{p}(^{27}$S$)=1230(450)$~keV.

\section{Discussion}
%\textit{Thermonuclear $^{26}$Si$(p,\gamma)^{27}$P reaction rates}
\subsection{Thermonuclear $^{26}$Si$(p,\gamma)^{27}$P reaction rate}
Considering the aforementioned analysis, the $^{26}$Si$(p,\gamma)^{27}$P reaction rate can be experimentally constrained. The Gamow windows for this reaction are calculated from a numerical study of the relevant energy ranges for astrophysical reaction rates~\cite{Rauscher_PRC2010}. At any given temperatures below 2.0~GK, the $3/2^+$ first resonance is always the closest one to Gamow peaks. The $5/2^+$ second and $5/2^+$ third resonances enter the Gamow window at temperatures above 1.2~GK and 2.0~GK, respectively. For the key $3/2^+$ resonance at 318(8)~keV, its proton partial width is calculated to be $\Gamma_p=2.55(74)$~meV using the relation $\Gamma_\gamma=\Gamma_p\times I_{\gamma}/I_p$, with the $\gamma$-ray partial width $\Gamma_\gamma=3.43(170)$~meV adopted in the compilation~\cite{Iliadis_NPA2010_3} and the experimental ratio of the $\gamma$-ray branch to the proton branch of $I_{\gamma}/I_p=1.35(39)$ determined in our work. Thus, an $\omega\gamma$ value of 2.92(191)~meV is obtained. Combining these values with the corresponding properties for the two $5/2^+$ resonances and the direct-capture component evaluated by Iliadis \textit{et al}.~\cite{Iliadis_NPA2010_3} yields the total rate based on the Monte Carlo techniques~\cite{Longland_NPA2010_1}, where uncertainties are rigorously defined.

\begin{figure}
\begin{center}
\includegraphics[width=7.5cm]{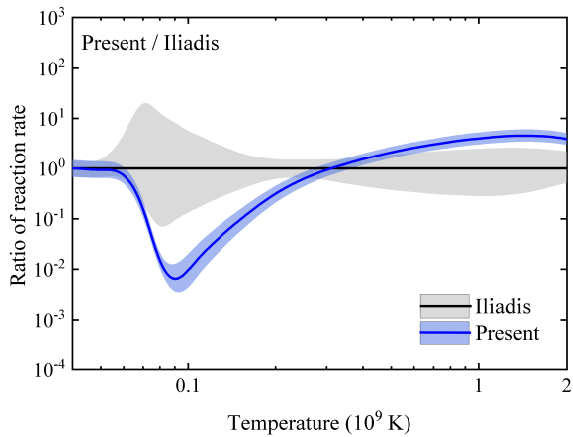}
\caption{\label{Reactionrate}Ratios of the $^{26}$Si$(p,\gamma)^{27}$P reaction rate determined in the present work to that evaluated by Iliadis \textit{et al}.~\cite{Iliadis_NPA2010_2}. The corresponding uncertainties are shown in shaded areas.}
\end{center}
\end{figure}

Currently, the $^{26}$Si$(p,\gamma)^{27}$P reaction rate from Iliadis \textit{et al}.~\cite{Iliadis_NPA2010_3,Iliadis_NPA2010_2} recommended in both REACLIB~\cite{Cyburt_APJS2010} and STARLIB~\cite{Sallaska_APJS2013} is universally adopted in astrophysical model calculations. As shown in Fig.~\ref{Reactionrate} the present rate is up to two orders of magnitude lower than the recommended rate in the temperature range $0.06<T<0.3$~GK typical for nucleosynthesis in nova. Our rate is higher than the recommended rate by up to a factor of 4 around 2.0~GK. The deviation is mainly caused by the larger resonance energy and strength for the $3/2^+$ resonance derived from our measurement. The present rate has much smaller uncertainties than the recommended rate almost over the entire temperature range, except that the present reaction rate follows the trend of the recommended rate below 0.06~GK where the direct-capture uncertainty dominates.
% The resonant rate can be derived from the resonance energies and strengths ($\omega\gamma$) of all isolated resonances that are located in the Gamow window~\cite{He_PRC2017,Lam_APJ2016,Rolfs_1988,Fowler_ARAA1967}, as the level density of the compound nucleus $^{27}$P is low.
% This accurate thermonuclear $^{26}$Si$(p,\gamma)^{27}$P rate will help determine the role of $^{26}$Si$(p,\gamma)^{27}$P reaction plays in the $rp$-process in type I x-ray bursts and novae explosions.

%\begin{table}\footnotesize
%\caption{\label{FirstRes27P}Resonance parameters of $3/2^+$ resonance in $^{26}$Si$(p,\gamma)^{27}$P reaction adopted in the present work and in literature.}
%\begin{center}
%\begin{tabular}{cccccc}
%Reference & $E_r$~(keV) & $\Gamma_\gamma$~(meV) & $\Gamma_p$~(meV) & $\Gamma_\gamma/\Gamma_p$  & $\omega\gamma$~(meV) \\
%  \hline
%\cite{Herndl_PRC1995} & 320   & 1.36  & 1.7   & 0.8 & 1.51  \\
%\cite{Caggiano_PRC2001} & 340(33) & 3.43  & 3.5   & 0.98 & 3.5  \\
%\cite{Guo_PRC2006} & 338 & 3.43 & 12.7(12) & 0.27  & 5.4(1) \\
%\cite{Togano_PRC2011} & 315(17) & 1.3 & 4.04(77)  & 0.32  & $1.9^{+1.9}_{-1.1}$ \\
%\cite{Marganiec_PRC2016} & 267(20) & 0.95 & 0.52 & 1.82 & 0.67 \\
%\cite{Janiak_PRC2017} & 332(30) &  &  & 1.3(2)$\sim$1.5(2) & \\
%\cite{Iliadis_NPA2010_3} & 259(28) & 3.4(17) & 0.180(72)  & 18.9(121) & 0.34(28) \\
%Present & 318(8) & 2.62(112) & 1.94(62)  & 1.35(39) & 2.23(134) \\
%Present & 318(8) & 0.91 & 0.67(20)  & 1.35(39) & 0.77(25) \\
%Present & 318(8) & 3.43(170) & 2.55(74)  & 1.35(39) & 2.92(191) \\
%\end{tabular}
%\end{center}
%\end{table}

%\textit{Implications for XRB model}
\subsection{Implications for XRB model}
%The computational demands on 3D hydrodynamics model simulations are substantial.
We have investigated the impact of present mass excesses of $^{27}$P and $^{27}$S on the composition of XRB nucleosynthesis zone using the one-zone post-processing nucleosynthesis code, which is a branch of the NuGrid framework~\cite{Herwig_PoS2009}, along with a trajectory K04 from Ref.~\cite{Koike_APJ2004}. It is shown that the nuclear energy generation remains unchanged in a process time of 100~s, but the abundances of $^{26}$Al and $^{26}$Si are increased by a factor of 2.4 in the last stage, compared to the abundances calculated using the recommended rate and mass from databases~\cite{Wang_CPC2017,Iliadis_NPA2010_2}. This is mainly attributed to the reverse $^{27}$P$(\gamma,p)^{26}$Si rate, which exponentially depends on the reaction $Q$-value. The higher $\mathrm{\Delta}(^{27}$P$)$ results in a significant increase in $^{27}$P$(\gamma,p)^{26}$Si rate, which will impede the proton capture process and leaves more $^{26}$Si and its corresponding $\beta$-decay daughter $^{26}$Al. Due to the neutron star gravitational potential, most of the burst ashes remain on the neutron star surface and replace the crust of the neutron star, and thus, they will have an impact on the accreted crusts thermal and compositional structure~\cite{Meisel_JPG2018}. A proper understanding of the ashes produced by XRBs is also important for the modeling of the crust evolution of accreting neutron stars~\cite{Schatz_N2014}. Similarly, the higher $\mathrm{\Delta}(^{27}$S$)$ would also result in a much stronger reverse $^{27}$S$(\gamma,p)^{26}$P rate which can effectively impact the final yield of $^{27}$S. The XRB model calculation shows that the final abundance ratio of $^{27}$S/$^{26}$P is 3500 using the $\mathrm{\Delta}(^{27}$S$)$ value from AME2016, compared to the $^{27}$S/$^{26}$P ratio of 0.4 if our $\mathrm{\Delta}(^{27}$S$)$ value is used. $^{27}$S was considered to be a waiting-point nucleus in the rapid proton capture process~\cite{Parikh_PRC2009,Parikh_PPNP2013}. However, this significant abundance change strongly implies that $^{27}$S cannot be regarded as a waiting-point nucleus anymore.
%at least at the level of 50\% over the whole duration of X-ray burst nucleosynthesis simulation.

%\textit{Implications for nova model}
\subsection{Implications for nova model}
In addition, the impact of our new $^{26}$Si$(p,\gamma)^{27}$P reaction rate on nova nucleosynthesis and in particular on the synthesis of $^{26}$Al has been examined through a series of hydrodynamic simulations. To this end, a suite of evolutionary sequences of nova outbursts hosting ONe white dwarfs of 1.15, 1.25, and 1.35~$M_\odot$ have been computed with the spherically symmetric, Lagrangian, hydrodynamic code SHIVA, extensively used in the modeling of novae and XRBs (see Refs.~\cite{Jose_2016,Jose_APJ1998} for details). Results have been compared with those obtained in three additional hydrodynamic simulations for the same white dwarf masses described above and the same physics inputs except for the $^{26}$Si$(p,\gamma)^{27}$P reaction rate, which was taken from the evaluation~\cite{Iliadis_NPA2010_2}. As confirmed by these simulations, the dominant destruction channel for $^{26}$Si in nova outbursts occurs via its $\beta^+$ decay to the isomeric state of $^{26}$Al, which subsequently decays to the ground state of $^{26}$Mg. No significant change in the element production in the Mg-P mass region was found when using the $^{26}$Si$(p,\gamma)^{27}$P reaction rate from Iliadis \textit{et al}.~\cite{Iliadis_NPA2010_2} or from the present work. Moreover, no significant changes were found when variations on this rate within uncertainties were used~\cite{Iliadis_APJS2002}. Compared to the result using the recommended Iliadis \textit{et al}.~\cite{Iliadis_NPA2010_2} rate, the contribution of classical nova outbursts to the galactic $^{26}$Al mass is only marginally increased by about 0.5\%. This verifies previous predictions of the nova contribution to the synthesis of galactic $^{26}$Al~\cite{Starrfield_PR1993,Jose_APJ1998} and places the expected $^{26}$Al/$^{27}$Al ratios in presolar grains of a inferred nova origin on solid experimental grounds~\cite{Jose_APJ2004}.

%It is expected that the mass uncertainty of $^{27}$P would affect the model predictions of XRB light curves strongly and also has a significant impact on the composition of the burst ashes~\cite{Schatz_APJ2017}, and the present result will provide a better constraint for modeling the nucleosynthesis in type I XRBs.

%\textit{Conclusion}
\section{Conclusion}
The $\beta$-decay spectroscopy of $^{27}$S yields the precise excitation energy of 1125(2)~keV, the proton-decay energy of 318(8)~keV, and the $\Gamma_\gamma/\Gamma_p$ ratio of 1.35(39) for the astrophysically important state in $^{27}$P. The proton-separation energy and mass excess of $^{27}$P are measured to a precision of 9~keV, a substantial improvement over previous measurements. In comparison with the recommended $^{26}$Si$(p,\gamma)^{27}$P reaction rate in REACLIB, the present rate is two orders of magnitude lower at temperatures around 0.1~GK, and the corresponding uncertainty in the rate is reduced by a factor of 9. A series of astrophysical model calculations incorporating the quantities obtained from this work were performed, and the dominant destruction mode of $^{26}$Si under nova temperatures is confirmed to be $\beta^+$ decay rather than the $^{26}$Si$(p,\gamma)^{27}$P reaction. We conclude that the new masses of $^{27}$P and $^{27}$S have no significant effects on the energy production in XRB, but the final abundances of $^{26}$Al and $^{26}$Si are both found to increase by a factor of 2.4 at the end of burst. The XRB model calculations with the new $^{27}$S mass also suggest that $^{27}$S is not a significant waiting point.
%The new rate is now sufficiently precise for modeling the observable features and nucleosynthesis in astrophysical explosive scenarios.
\section{Acknowledgements}
We acknowledge the dedicated effort of the HIRFL beam physicists and operations staff for providing high-quality beams. We gratefully acknowledge Christian Iliadis for the reaction rate calculations. We would like to thank Zhihong Li, Bing Guo, Christopher Wrede, and Hendrik Schatz for very helpful discussions. This work is supported by the Ministry of Science and Technology of China under the National Key R\&D Programs Nos. 2018YFA0404404 and 2016YFA0400503, and by the National Natural Science Foundation of China under Grants Nos. 11805120,  11805280, 11825504, 11811530071, 11705244, 11705285, 11775316, U1732145, 11635015, 11675229, U1632136, 11505293, 11475263, 11490562, and U1432246, and by the Youth Innovation Promotion Association of Chinese Academy of Sciences under Grant No. 2019406, and by the China Postdoctoral Science Foundation under Grants Nos. 2017M621442 and 2017M621035, and by the Office of China Postdoctoral Council under the International Postdoctoral Exchange Fellowship Program (Talent-Dispatch Program) No. 20180068.% L. J. Sun, X. X. Xu, and S. Q. Hou contributed equally to this work and should be considered as co-first authors.

\end{document}